\documentclass{mem}
\usepackage{natbib}\usepackage{txfonts}\usepackage{balance}
\usepackage{graphicx}
\usepackage[a4paper,breaklinks,dvipdfm]{hyperref}
\idline{75}{282}
\begin{document}
\def\teff{$T\rm_{eff }$}
\def\kms{$\mathrm {km s}^{-1}$}

\title{Chemical evolution in star clusters: the role of mass and environment}

   \subtitle{}

\author{S. L. \,Martell\inst{1} \and S. \,Duffau\inst{2} \and A. P. \,Milone\inst{3,4} \and G. H. \,Smith\inst{5} \and M. M. \,Briley\inst{6} \and E. K. \,Grebel\inst{7}}

\institute{
Australian Astronomical Observatory, PO Box 915, North Ryde NSW 1670, Australia\\
\email{smartell@aao.gov.au}
\and
Landessternwarte K\"{o}nigstuhl, Zentrum f\"{u}r Astronomie der Universit\"{a}t Heidelberg, 69117 Heidelberg, Germany
\and 
Instituto de Astrofisico de Canarias/Department of Astrophysics, University of La Laguna, E-38200 La Laguna, Tenerife, Canary Islands, Spain
\and
Research School of Astronomy and Astrophysics, Australian National University \& Mount Stromlo Observatory, Cotter Road, Weston, ACT, 2611, Australia
\and
University of California Observatories/Lick Observatory, Department of Astronomy \& Astrophysics, UC Santa Cruz, Santa Cruz, CA 95064, USA
\and
Department of Physics and Astronomy, Appalachian State University, Boone, NC 28608, USA
\and 
Astronomisches Rechen-Institut, Zentrum f\"{u}r Astronomie der Universit\"{a}t Heidelberg, 69120 Heidelberg, Germany
}

\authorrunning{Martell et al}

\titlerunning{LMC clusters}

\abstract{The process of chemical self-enrichment in stellar systems can be affected by the total mass of the system and the conditions of the large-scale environment. Globular clusters are a special dark matter-free case of chemical evolution, in which the only self-enrichment comes from material processed in stars, and only two bursts of star formation occur. We describe how observations of intermediate-age star clusters in the Large Magellanic Cloud can provide insight on the ways that mass and environment can affect the process of chemical enrichment in star clusters.
}
\maketitle{}

\section{Introduction}
Although they have long been held up as examples of simple stellar systems, all Galactic globular clusters host significant complexity in light-element abundances (e.g., Carretta et al. 2009\nocite{CB09}; Lardo et al. 2011\nocite{LB11}; Smolinski et al. 2011\nocite{SM11}), and some also have variations in alpha-, iron-peak and neutron-capture element abundances (e.g., Cohen \& Kirby 2012\nocite{CK12}; Carretta et al. 2010\nocite{CG10}; Marino et al. 2011\nocite{MS11}). The light-element abundance complexity, and the corresponding dramatic photometric complexity (e.g., Piotto et al. 2012\nocite{PM12}; Milone et al. 2012\nocite{MP12}), are presently explained as a result of a constrained chemical feedback between two closely-spaced stellar generations (e.g., D'Ercole 2008\nocite{DVD08}; Conroy \& Spergel 2011\nocite{CS11}). In this ``stellar-mode'' self-enrichment process, material that has been processed through the CN(O), Ne-Na and Mg-Al cycles of hydrogen fusion is ejected in the winds of massive main-sequence stars and AGB stars, and during lossy mass transfer in close binaries (e.g., Decressin et al. 2007\nocite{DMC07}; D'Ercole et al. 2008\nocite{DVD08}; de Mink et al. 2009\nocite{DPL09}). This material mixes with pristine material left over from the formation of the initial population of stars to form a second generation with light-element abundances that fall between the scaled-Solar abundances of the initial population and the highly anticorrelated fusion-processed abundance pattern of the feedback material. Because this process happens quickly, there is not time for supernova winds to contribute enhancements in either alpha-element or iron-peak abundances; indeed, the large amount of kinetic energy deposited by supernova winds clears all remaining gas from the cluster and drives the cluster to expand, causing cluster stars at large radius to become gravitationally unbound.

The stellar-mode self-enrichment process that occurs in globular clusters can be thought of as the low-mass end of a continuum of chemical evolution processes, continuing into ultrafaint dwarf galaxies, which have sporadic star formation that captures snapshots of the galay's chemical state (e.g., Gilmore et al. 2012\nocite{GN12}); classical dwarf galaxies, which typically have bursty star formation but a smooth progression from SN II-driven chemical evolution to SN Ia-driven chemical evolution (e.g., Koch et al. 2008\nocite{KG08}; Revaz et al. 2009\nocite{RJ09}; Kirby et al. 2011\nocite{KL11}); and galaxies at or above the mass of the Milky Way, in which the overall mass, luminosty and metallicity follow well-defined correlations. Stellar feedback processes must contribute to the interstellar medium in all of these systems, but it is only in globular clusters that it produces a noticeable abundance effect: presumably the signature of hot hydrogen fusion is washed out by the dramatically larger mass in heavier elements produced by supernovae.

\section{A minimum mass for self-enrichment?}
The ability of a dark matter-free system like a globular cluster to retain stellar winds is necessarily related to its mass at the time of star formation, with some minimum mass below which star clusters, though their stars produce fusion-processed winds, are not able to convert those winds into new stars. This raises the question of why there are no long-lived, chemically simple star clusters in the Milky Way: were they never formed, or is the minimum mass for self-enrichment quite low, or is the Galaxy such an inhospitable place for low-mass star clusters that any single-generation star clusters that were initially formed have since been destroyed by tidal forces, disk shocking and internal 2-body interactions (e.g., Gnedin \& Ostriker 1997\nocite{GO97})? Efforts to reconstruct the initial cluster mass function based on the present-day log-normal globular cluster mass function and various cluster-dissolution processes (e.g., Fall \& Zhang 2001\nocite{FZ01}; Parmentier \& Gilmore 2007\nocite{PG07}; Baumgardt et al. 2008\nocite{BK08}) do not find consistent results.

The initial masses of globular clusters are not directly observable, but attempts have been made to correlate present-day masses to the presence of light-element abundance complexity (e.g., Kayser et al. 2008\nocite{KH08}; Carretta et al. 2010a\nocite{CB10}). In \citet{CB10} it is pointed out that all Galactic globular clusters with masses above a few $\times 10^{\rm 4} M_{\odot}$ that have been surveyed spectroscopically have been found to host multiple light-element abundance populations, with a wider abundance range found in the more massive clusters. We consider here an environmental explanation for the lack of single-generation globular clusters in the Milky Way. This explanation leaves open the possibility that other environments might be more hospitable to the long-term survival of lower-mass clusters.

\begin{figure}[]
\resizebox{\hsize}{!}{\includegraphics[clip=true]{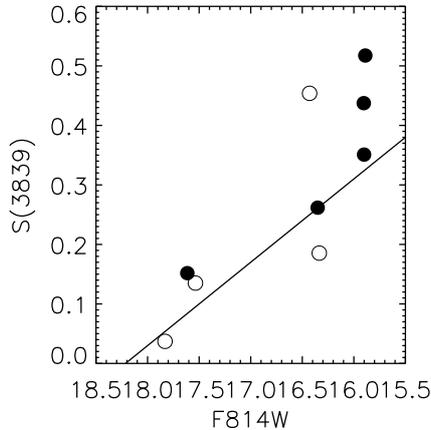}}
\caption{
\footnotesize
CN band strength index $S(3839)$ versus apparent F814W magnitude for RGB members of NGC 1651 (open circles) and NGC 1751 (filled circles). An index baseline is shown as a solid line, following the trend of the lower edge of the data as described in \citet{N81}.
}
\end{figure}

\section{Star clusters in the Large Magellanic Cloud}
As a way to explore the influence of the large-scale environment on chemical complexity in star clusters, we have taken moderate-resolution spectra of individual red giant stars in the intermediate-age populous LMC star clusters NGC 1651 and NGC 1751. Using the FORS2 spectrograph \citep{S94} on UT1 (Antu) at the VLT, we obtained spectra for 4 RGB stars belonging to NGC 1651 and 5 in NGC 1751 on 17 December 2011 as part of ESO program 088.D-0807. From the literature, it was unclear what result we should expect: although \citet{MO09} found light-element abundance variations among RGB stars in old globular clusters in the LMC, \citet{MC08} found only a narrow O-Na anticorrelation for RGB stars in intermediate-age LMC clusters. Figure 1 shows the CN band strength index $S(3839)$, measured from our FORS2 spectra following the definition of \citet{N81}, versus apparent {\it HST} ACS/WFC F814W magnitude for all of these stars, with NGC 1651 stars shown as open circles and NGC 1751 stars shown as filled circles. In both clusters, there is a distinct range in $S(3839)$ at fixed luminosity, indicating that stars in both clusters exhibit a range in nitrogen abundance. 

Figure 2 shows a generalized histogram of $\delta S(3839)$ (the vertical distance between each star's CN band strength and the baseline shown in Fig. 1) for both clusters together. Since they have very similar metallicities and ages ([Fe/H]$=-0.3$, age$=1.8$ Gyr for NGC 1651 and [Fe/H]$=-0.44$, age$=1.5$ Gyr for NGC 1751), data for the two clusters can be combined this way without significant differential effects on the CN band strengths. In this figure there is a distinct width to the distribution, indicating a range of nitrogen abundance in the stars.

\begin{figure}[]
\resizebox{\hsize}{!}{\includegraphics[clip=true]{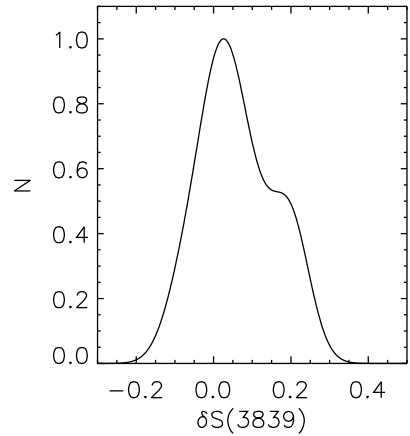}}
\caption{
\footnotesize
Generalized histogram of the distribution of $\delta S(3839)$ for the 9 RGB members of NGC 1651 and NGC 1751, composed of an individual Gaussian for each star with a FWHM of 0.05 dex, the typical measurement error on $\delta S(3839)$. 
}
\end{figure}

This is particularly interesting in light of the recent photometric work by \citet{MB08}, \citet{MB09}, and Goudfrooij et al. (2011a, 2011b)\nocite{GP11a}\nocite{GP11b}, which found that the main sequence turnoffs in a significant fraction of populous intermediate-age LMC clusters are broadened or split. This has been interpreted as a sign of extended or two-burst star formation in those clusters. In this context, we view our CN band strength data as a sign that the complex star formation histories in intermediate-age populous star clusters in the LMC included a phase of stellar-mode self-enrichment, though further work is required to solidify this claim.

It has been suggested by \citet{CS11} that star clusters moving in the gravitational potential of the LMC will be more able to accrete material from their surroundings than star clusters belonging to the Milky Way because of their lower orbital velocities. This would then lower the effective minimum mass for the formation of a second stellar generation, since (setting aside the problem of maintaining a consistent metallicity) accretion would allow gas from all along the cluster's orbit, and not just the gas initially associated with the cluster, to contribute to the construction of the second generation.

\section{Conclusions}
We find evidence for complexity in the light-element abundances of NGC 1651 and NGC 1751. These two clusters demonstrate that the multiple stellar generations suggested by the broadened or split main sequences in intermediate-age LMC clusters can be accompanied by complex abundances. Both are at or above the minimum present-day mass for self-enrichment for Galactic globular clusters \citep{BP12}, which unfortunately does not provide any information about whether the minimum mass for self-enrichment is lower in the LMC environment. However, the presence of complex abundances in intermediate-age clusters does support the claim of \citet{KM11} that multi-generation star formation may be a standard feature of star cluster formation, and is not restricted to ancient globular clusters.

\bibliographystyle{aa}


\end{document}